# Chip-Scale, Sub-Hz Fundamental Sub-kHz Integral Linewidth 780 nm Laser through Self-Injection-Locking a Fabry-Pérot laser to an Ultra-High Q Integrated Resonator

Andrei Isichenko[1], Nitesh Chauhan[1], Kaikai Liu[1], Mark Harrington[1], and Daniel J. Blumenthal[1*]

[1] Department of Electrical and Computer Engineering, University of California Santa Barbara, Santa Barbara, CA 93106 USA

* Corresponding author (danb@ucsb.edu)

## ABSTRACT

Today's state of the art precision experiments in quantum, gravimetry, navigation, time keeping, and fundamental science have strict requirements on the level and spectral distribution of laser frequency noise. For example, the laser interaction with atoms and qubits requires ultra-low frequency noise at multiple offset frequencies due to hyperfine atomic transitions, motional sidebands, and fast pulse sequencing. Chip-scale integration of lasers that meet these requirements is essential for reliability, low-cost, and weight. Here, we demonstrate a significant advancement in atomic precision light sources by realizing a chip-scale, low-cost, 780 nm laser for rubidium atom applications with record-low 640 mHz (white noise floor at 0.2 Hz$^2$/Hz) fundamental and 732 Hz integral linewidths and a frequency noise that is multiple orders of magnitude lower than previous hybrid and heterogeneous self-injection locked 780 nm lasers and lower noise than bulk microresonator implementations. The laser is a Fabry-Pérot laser diode self-injection locked to an ultra-high Q photonic integrated silicon nitride resonator. This performance is enabled by a 145 million resonator Q with a 30 dB extinction ratio, the highest Q at 780 nm, to the best of our knowledge. We analyze the impact of our frequency noise on specific atomic applications including atomic frequency references, Rydberg quantum gates, and cold atom gravimeters. The photonic integrated resonator is fabricated using a CMOS foundry-compatible, wafer-scale process, with demonstrated integration of other components showing promise for a full system-on-a-chip. This performance is scalable to other visible atomic wavelengths, opening the door to a variety of transitions across many atomic species and enabling low-power, compact, ultra-low noise lasers impacting applications including quantum sensing, computing, clocks and more.

Chip-Scale, Sub-Hz Fundamental and Sub-kHz Integral Linewidth 780 nm Fabry-Pérot Laser through Self-Injection-Locking using an Ultra-High Q Integrated Resonator

# INTRODUCTION

Integrated ultra-narrow linewidth visible and near-IR emission lasers are critical for the miniaturization, improved reliability, and scaling of atomic systems such as quantum computing [1], precision sensing [2], and timekeeping [3]. These applications have stringent requirements on the spectral properties of the laser frequency noise over several decades of frequency offset from carrier. For systems involving manipulating and interrogating atoms and qubits, evaluating the probing laser frequency noise requires a consideration of neighboring hyperfine atomic transitions and motional sidebands as well as Fourier components generated by pulse sequencing and other control signals applied to the laser and atom. For example, ultra-narrow linewidth 780 nm lasers at the rubidium atom wavelength are key for applications including 2-photon atomic clocks [4], cold atom interferometer sensors [2], and neutral atom quantum information processing [1]. Traditionally, these systems employ costly external cavity lasers [5] and today these lasers continue to be relegated to costly, bulky table-top systems and often utilize frequency doubling of mid-IR lasers. Photonic integration is critical to realizing compact, lightweight, low cost and portable atomic and quantum systems. Yet ultra-narrow linewidth visible and near-IR lasers that leverage integration technologies have remained elusive. An important approach is direct-drive (direct emission without frequency conversion) optical self-injection locking (SIL) [6]. Bulk-optic 780 nm SIL lasers achieve Hz-level fundamental linewidths [7]. The next step in quantum and atomic systems is realization of direct-drive high performance visible and near-IR lasers in a photonic integrated, CMOS foundry-compatible platform. This milestone will impact a wide range of applications including portable atomic clocks [8] and space-based quantum sensors [9].

Bulk optic crystalline ultra-high Q WGMRs have been used to realize 780 nm SIL lasers with fundamental linewidth of 5 Hz and a $1/\pi$ integral linewidth of 1.4 kHz [7]. However, WGMRs pose a challenge for systems-on-chip integration and wafer-scale foundry processing. Integration using CMOS foundry compatible processes such as the ultra-low loss silicon nitride (SiN) [10] is key to reducing cost and weight, improving robustness to environmental disturbances, and integration of atomic and quantum systems-on-chip. However, any integration solution in visible and near-IR must address the strict requirements on laser frequency noise distribution for atomic and quantum systems, which to date has remained elusive. Today, at 1550 nm, exquisite fundamental and integral laser noise has been achieved with integrated high-Q SiN resonators that have reduced fundamental linewidths to 88 mHz [11] and integral linewidths to 36 Hz [12]. Yet at 780 nm, corresponding to the important $D_2$ transition in rubidium, integrated SIL Fabry-Pérot (FP) laser diodes butt-coupled to 80,000 Q, 10 dB extinction ratio (ER) SiN resonators have only yielded 700 Hz fundamental and 50 kHz integral linewidths [13]. Compact 780 nm SIL lasers locked to narrow band filters have demonstrated 10s of kHz beat linewidth [14] and intra-cavity SIL frequency-doubled lasers achieved 12 Hz fundamental linewidth with 30 $\mu$W output power [15]. Heterogeneous laser integration is an important approach and recently a 780 nm heterogeneous GaAs laser butt-coupled to a 10 M intrinsic Q SiN ring was used to reach 92 Hz fundamental linewidth [16], with no integral linewidth reported. To date, the lowest reported loss at 780 nm is 0.1 dB/m in a silica wedge resonator [17]. The lowest reported visible losses and highest Q for fully integrated resonators are 0.65 dB/m loss and 90 million intrinsic Q at 674 nm [18] and low linewidth integrated visible light lasers include a 674 nm stimulated Brillouin laser (SBL) achieve a 269 Hz fundamental linewidth [19]. To reach the ultra-low frequency noise regime at 780 nm, new direct-drive laser technologies based on ultra-low loss waveguides and ultra-high Q, large mode volume resonators are needed.

We demonstrate a significant advance in direct-drive atomic precision light sources by realizing a sub-Hz fundamental, sub-kHz integral linewidth 780 nm laser for rubidium in a 200 mm CMOS foundry compatible, wafer-scale, silicon nitride integration platform. The extremely high Q and ER enables SIL of a low-cost FP laser diode, reducing the free running linewidth from many GHz to 732 Hz integral linewidth, and the fundamental linewidth to 640 mHz (white noise floor at 0.2 Hz$^2$/Hz). This level of performance is possible due to the remarkable 0.36 dB/m waveguide loss, 145 million intrinsic resonator Q, and 30 dB extinction ratio (ER). This is the lowest loss and highest Q at 780 nm to date to the best of our knowledge, reaching thermorefractive noise (TRN) limited performance that is enabled by the moderate confinement, large mode volume resonator. We achieve performance rivaling that of bulk-optic resonators with an integrated chip with a simple design and low cost, eliminating the need to complex fiber-to-resonator coupling and other micro-optics. The platform performance is scalable to other visible atomic wavelengths [20] opening the door to a wide variety of transitions across many atomic species. The exceptional noise performance is achieved due to the strong optical feedback, which is powerful enough to allow for extracting the laser output with an intra-cavity tap while maintaining SIL. These results represent orders of magnitude improvement in fundamental and integral linewidths over previous 780 nm SIL demonstrations. The CMOS foundry platform and process are fully compatible with other passive and active components [10,21], showing promise for full systems-on-chip integration.

## RESULTS AND DISCUSSION

We demonstrate ultra-low frequency noise injection locking using a commercial 780 nm FP laser diode coupled to an integrated ultra-high Q, high extinction ration (ER) Si$_3$N$_4$ resonator. The coupling is simple and low-cost, demonstrating the advantage of using this high-performance silicon nitride technology and integrated resonator. This exquisite performance is achieved with the laser coupled using a simple 3-meter fiber pigtail and fiber splitter (Fig. 2a). The strength of the feedback provided by the ultra-high Q resonator makes this laser robust to fluctuations in the fiber and splitter, which would not be possible using other lower Q and higher loss technologies. While the subject of hybrid and heterogeneous packaged lasers to resonators is not the focus in this work, it is a straightforward extension using techniques already reported for visible and near-IR integrated lasers [16,22,23]. In the future, the laser, tap, and passive components can be integrated directly with the resonator [24] and this additional packaging will further improve the locking stability and noise performance beyond that reported here.

The resonators are fabricated using a 200 mm CMOS foundry-compatible process [10] (Fig. 1a). The Si$_3$N$_4$ waveguides consist of a 15 µm SiO$_2$ lower cladding, a 40 nm thick and 4 µm wide Si$_3$N$_4$ core, and a 6 µm SiO$_2$ upper cladding (Fig. 1b) and support fundamental TE and TM modes, with the TM mode chosen due to its lower loss [25]. An intrinsic $Q_i$ = 145 M and loaded $Q_L$ = 65 M are measured using an unbalanced Mach-Zehnder interferometer (MZI) [26] (Fig. 1c). The 0.36 dB/m propagation loss is calculated by fitting two resonators of the same radius (5.84 mm) with different coupling coefficients. The ultra-high Q and a 30 dB ER enable strong optical feedback through Rayleigh backscattering inside the resonator cavity [7]. The laser diode is a Thorlabs LPS-785-FC with no internal isolator. The resonator is placed on a temperature-controlled copper block and the cleaved input and output fibers at the bus waveguide are positioned using standard 3-axis stages. SIL is verified by monitoring the resonator transmission as the laser current is ramped (Fig. 2b) and is demonstrated by the characteristic vertical edge features of the transmission signal [27].

SIL of the low-cost FP laser to the photonic integrated resonator results in single mode operation with many orders magnitude reduction in the FP laser linewidth. We achieve a fundamental linewidth that is 8 times lower than that of a previous micro-resonator bulk-optic demonstration [7], over two orders of magnitude lower for direct-emission integrated 780 nm resonators [13,16], and frequency noise at certain frequencies that is over 3 orders of magnitude lower than in recently reported hybrid-integrated SIL [13] and heterogeneous integrated SIL [16]. The frequency noise (FN) is measured using an optical frequency discriminator (OFD) [19] and reaches a minimum value of 0.2 $Hz^2$/Hz corresponding to a fundamental linewidth of 640 mHz and a 732 Hz $1/\pi$ integral linewidth [12] (Fig. 3). The actual integral linewidth is likely lower as the noise below 1 kHz frequency is dominated by the OFD measurement noise [22]. A 1.5 mW output power is measured after the fiber splitter tap located inside the cavity (Fig. 2a). The FN is measured with the PIC uncoupled and although back-reflections from fiber components promote single-mode operation of the laser, the frequency noise does not reach that of the PIC SIL. We also confirm that the laser is on resonance by monitoring the resonator transmission port (Fig. 1c). Importantly, the frequency noise reaches the simulated resonator TRN limit [28] (dashed black curve, Fig. 3) which is low due to the large optical mode volume [12].

The development of integrated, narrow-linewidth laser sources is critical for atomic and quantum systems and requires careful consideration of the frequency noise (FN) requirements associated with various applications. The performance of these systems is influenced by the spectral distribution of the incident optical local oscillator (LO) FN which affects the transition probabilities between atomic energy levels. Understanding the impact of FN, rather than stating a total linewidth, is essential because different applications have distinct sensitivity to noise at different frequency bands. Low frequency drift can cause the laser to deviate from the atomic transition absolute frequency. Mid-frequency (offset from carrier) noise can dominate the residual lock loop noise, and high frequency noise can lead to cross-talk with neighboring hyperfine transitions and motional sidebands. In systems that are periodically interrogated by the LO, particularly with fast pulses, noise at high frequencies can alias into the readout of the atomic or quantum sensor. To quantify the effect of laser FN on overall performance, an application-specific filter function can be employed to weight of the importance of frequency components in the laser FN distribution.

The performance of rubidium atom sensor and computing technologies can be directly affected by the noise of narrow-linewidth lasers. In Table 1 we summarize the laser FN contribution for our 780 nm SIL to the performance of several rubidium atomic applications. The methods used to develop Table 1 are applicable to many other wavelengths, transitions, and applications. We compare the performance of our laser to that of a typical free-running semiconductor (FRSC) laser as well as that of recently published integrated results [13,15,16] and discuss how laser FN limits the performance of three distinct atomic systems that could benefit from photonic integration for future scalability and portability. The process of optically interrogating an atomic transition introduces a time-varying sensitivity to the local oscillator (LO) laser FN [29]. In the case of atomic frequency standards such as a two-photon rubidium optical clocks, this effect is known as intermodulation noise [30]. The short-term clock instability scales with the LO FN at twice the modulation frequency used in the stabilization [31]. We calculate that our reported ultra-low FN SIL laser, if used as a probe for a two-photon atomic frequency reference, can achieve a short-term ($\tau \approx 1$ s) stability of $2.6 \times 10^{-15}$ at 1 s, almost 10x lower than that of a previously reported FN-limited compact optical standard [31], when intermodulation noise is the dominant limiting factor. For quantum computing with Rydberg atoms, laser pulses containing excess FN can impact quantum gate fidelities. High offset frequency FN results in Rabi oscillation dampening and limits the achievable lifetime of Rydberg states [32]. Active laser stabilization to bulk reference cavities can introduce feedback loop noise at MHz frequency offsets, and filtering this noise has been shown to increase Rydberg state coherence times from 7 $\mu$s to over 20 $\mu$s [33]. A similar pulsed interrogation scheme can cause sensitivity limits in cold atom interferometer gravimeters used for gravity surveys [2,34]. The probe laser's FN affects gravity measurement sensitivity through a transfer function $H(f)$ with parameters related to the cold atom experiment such as sensor bandwidth and Raman laser pulse duration [34]. We calculate the gravity measurement sensitivity limit assuming that the instrument is limited by the Raman laser frequency noise. In general, gravimeter measurements have required lower-noise, bulky ECDL lasers because FRSC lasers

Chip-Scale, Sub-Hz Fundamental and Sub-kHz Integral Linewidth 780 nm Fabry-Pérot Laser through Self-Injection-Locking using an Ultra-High Q Integrated Resonator with ~1 MHz integral linewidths may degrade the gravimeter sensitivity [34], which may exceed noise contributions normally dominated by sensor vibration noise.

The integration of foundry-compatible photonics offers significant potential to enable fully integrated atomic and quantum systems at the chip-scale, improving scalability and robustness. Our ultra-low-loss platform that enables these record-low frequency noise results has also been demonstrated in other relevant chip-scale functionalities. Importantly, the requirements of these applications extend beyond achieving low frequency noise; precise control over the absolute laser frequency is also essential. In this regard, our ultra-high Q resonator can be metallized with a thermal tuner such that the ultra-low optical loss is maintained, enabling tuning a locked laser for stabilization to atomic spectroscopy [35]. Advancements such as faster modulation with PZT-on-SiN [21] have also been demonstrated without affecting optical loss, presenting opportunities for pulse generation and implementing stabilization loops for atomic systems-on-chip. Furthermore, SiN photonic integrated beam delivery in a cold atom magneto-optical trap has achieved large output beam diameters, enabling successful trapping of over a million rubidium atoms [36]. This atom number is promising for surpassing shot noise limits in sensors such as atom interferometers. These results demonstrate the potential of integrating an ultra-low linewidth SIL laser as a probe for compact atomic sensors, clocks, and a wide range of precision atomic and quantum technologies.

**Table 1.** Atomic and quantum laser frequency noise requirements and performance limits and comparison between this work and free-running semiconductor lasers (FRSC).

| Application | Offset frequency | Frequency noise at offset (Hz²/Hz) — This work | Limiting noise source[1] | Laser FN-limited performance — FRSC / [13] / [15] / [16] | Laser FN-limited performance — This work |
|---|---|---|---|---|---|
| Cold atom interferometer gravimeter [34] | 24 kHz 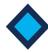 | 12 | Gravity sensitivity limited by Raman laser frequency noise[3] $$\sigma_g^2 \sim \int_0^\infty |H(f)|^2 \, S_{\Delta\nu}(f) \, df$$ | 105 / 43 / 23 / 46 nm/s²/$\sqrt{\mathrm{Hz}}$ | 2 nm/s²/$\sqrt{\mathrm{Hz}}$ |
| 2-photon atomic wavelength reference at $f_0$ for 778 nm laser [4,30] | 160 kHz, $f_m = 80$ kHz 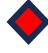 | 4 | Short-term ($\tau \approx 1$ s) instability due to intermodulation (IM) effect [30,31] $$\sigma_y^{(IM)}(\tau) = \frac{[S_{\Delta\nu}(2f_m)/f_0^2]^{1/2}}{2\sqrt{\tau}}$$ | $1.4 \times 10^{-12}$ / $3.9 \times 10^{-14}$ / $1.9 \times 10^{-14}$ / $7 \times 10^{-14}$ | $2.6 \times 10^{-15}$ |
| High-fidelity Rydberg gates [33] | 1 MHz 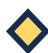 | 0.5 | Rabi oscillation decay time due to laser FN at high offset frequencies [32] | 3 / 21 / 22 / 22 μs | 22 μs |

[1] Noise contribution related laser frequency noise, assuming all other noise sources are not considered in evaluating the system performance.
[2] Free-running semiconductor (FRSC) laser such as a distributed Bragg reflector (DBR) laser, for comparison.
[3] Using cold atom interferometer parameters from [34].

Chip-Scale, Sub-Hz Fundamental and Sub-kHz Integral Linewidth 780 nm Fabry-Pérot Laser through Self-Injection-Locking using an Ultra-High Q Integrated Resonator

# FIGURES

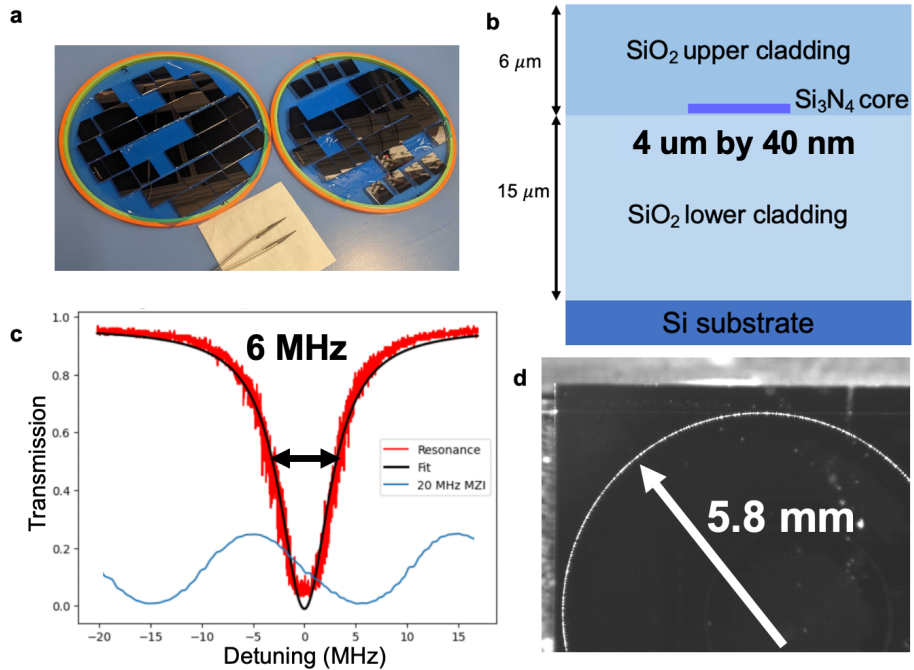

**Fig. 1. Ultra-high Q resonator at 780 nm.** a) 780 nm resonators fabricated in a 200 mm CMOS foundry. b) Device waveguide cross-section. c) Q measurement of resonator at 780 nm. d) Infra-red camera image of the laser locked to the ring resonance.

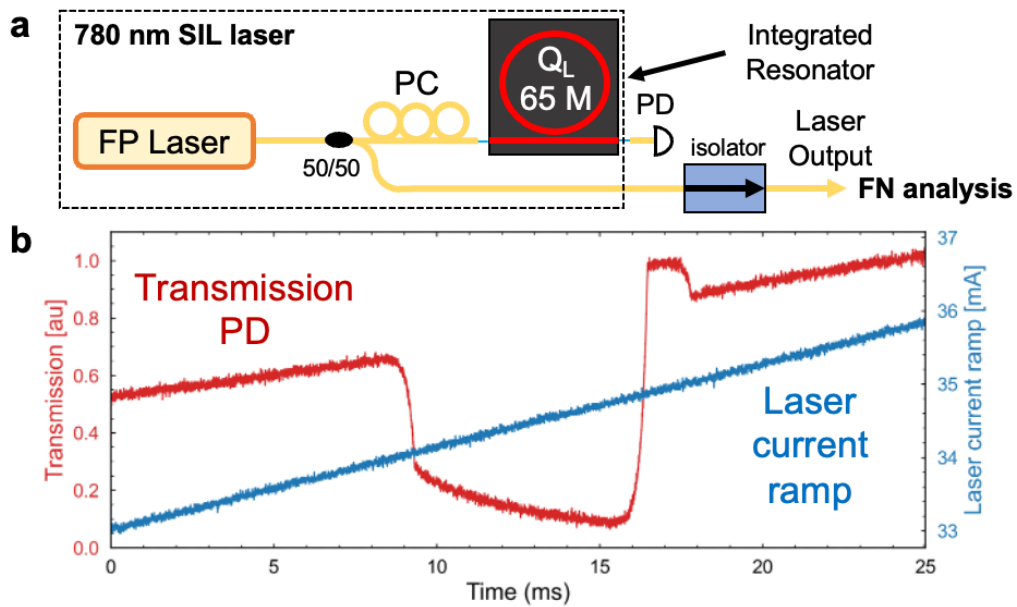

**Fig. 2. SIL system.** a) Schematic of the SIL setup. The SIL laser output is fed into the OFD measurement setup for frequency noise measurements. PC: polarization controller.  b) Laser current ramp and monitoring the transmitted power.

Chip-Scale, Sub-Hz Fundamental and Sub-kHz Integral Linewidth 780 nm Fabry-Pérot Laser through Self-Injection-Locking using an Ultra-High Q Integrated Resonator

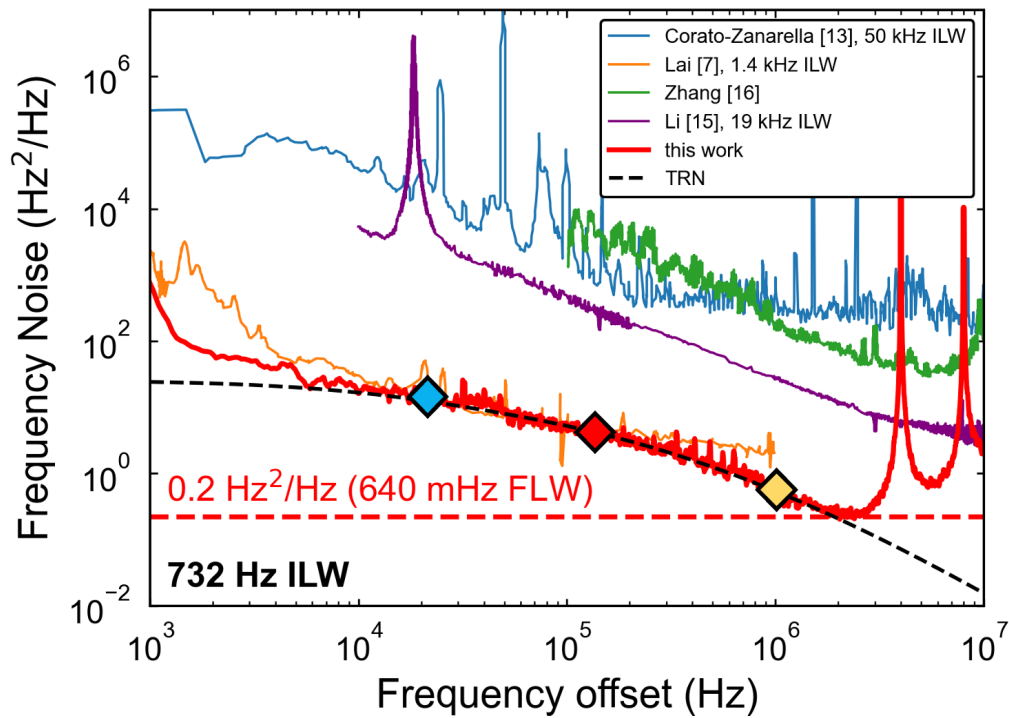

**Fig. 3. Frequency noise results.** OFD frequency noise data of the self-injection locked FP laser and the simulated TRN noise floor. The integral linewidth (ILW) calculated with the reverse integration $1/\pi$ method yields 732 Hz. Spurs at multiples of 4 MHz correspond to the free-spectral range of the OFD MZI and do not contribute to the integral linewidth calculation. Diamonds corresponds to relevant frequency offsets discussed in Table 1: 24 kHz (blue), 160 kHz (red), 1 MHz (yellow).

# FUNDING

This material is based upon work supported by Infleqtion/ColdQuanta and a UCSB Faculty Research Grant.

# ACKNOWLEDGEMENTS

We thank Karl Nelson at Honeywell for the fabrication. We acknowledge Andrey Matsko at NASA Jet Propulsion Laboratory for useful discussions and Ryan Behunin at Northern Arizona University for help with thermorefractive noise simulations.

# DISCLOSURES

The authors declare no conflicts of interest.

# DATA AVAILABILITY

Data underlying the results presented in this paper are not publicly available at this time but may be obtained from the authors upon reasonable request.

Chip-Scale, Sub-Hz Fundamental and Sub-kHz Integral Linewidth 780 nm Fabry-Pérot Laser through Self-Injection-Locking using an Ultra-High Q Integrated Resonator